\documentstyle[12pt]{article}
\advance\textheight by \topskip
\begin{document}
\thispagestyle{empty}
\begin{flushright}
{SU-ITP-94-39}\\
hep-th/9411111\\
November 15, 1994
\end{flushright}
\vskip 1.2cm
\begin{center}
{\Large\bf DO WE LIVE IN THE CENTER OF THE WORLD?}
\vskip 1.3cm
{\bf Andrei Linde}\footnote{E-mail: linde@physics.stanford.edu}\\
\vskip  0.2cm
Department of Physics, Stanford University, Stanford CA 94305-4060,
 USA
\vskip 0.3cm
{\bf Dmitri Linde}\footnote{E-mail: dmitri@cco.caltech.edu}\\
\vskip 0.2cm
California Institute of Technology, Pasadena, CA 91125,
 USA
 \vskip 0.15cm
and
\vskip 0.2 cm
{\bf Arthur Mezhlumian}\footnote{On leave
from: Landau
 Institute for Theoretical Physics, Moscow. \
 E-mail: arthur@physics.stanford.edu}\\
 \vskip 0.2cm
Department of Physics, Stanford
University, Stanford CA 94305-4060, USA\\
\end{center}

\vskip .5 cm
{\centerline{\large ABSTRACT}}
\begin{quotation}
\vskip -0.3cm
We investigate the distribution of energy density in a stationary
self-reproducing inflationary universe. We show that  the main fraction of
volume of the  universe in a state with a given density $\rho$ at any given
moment of time $t$ in synchronous coordinates is concentrated near the centers
of deep exponentially wide spherically symmetric holes in the density
distribution.
A possible interpretation of this result is that a typical observer should see
himself living in the center of the world. Validity of this interpretation
depends  on the choice of measure in quantum cosmology. Our investigation
suggests that unexpected (from the point of view of inflation) observational
data, such as possible local deviations from $\Omega = 1$, or possible
dependence of the Hubble constant on the length scale, may tell us something
important about quantum cosmology and  particle physics at nearly Planckian
densities.

\end{quotation}
 \newpage

\noindent1.\hskip 0.4 cm    In this paper we are going to show   that according
to a very wide class of inflationary theories,   the main fraction of volume of
the universe  in a state with a given density $\rho$ at any given  moment of
time $t$ (during or after inflation) should be concentrated near the centers of
deep exponentially wide spherically symmetric holes in the density
distribution.

Observational implications of this results depend  on its interpretation.  If
we assume that we live in a part of the universe which is typical, and by
``typical'' we mean  those parts of the universe which have the greatest volume
with  other parameters (time and density) being equal, then our result  implies
 that we should live near the center of one of the holes in the density
distribution. There should be many such holes in the universe, but each of them
  should be exponentially wide.  Therefore an  observer living near the center
of any such hole will see himself  ``in the center of  the world''.

One should clearly  distinguish between the validity of our result and the
validity of its interpretation suggested above. Even though the effect by
itself is rather surprising (its existence was first conjectured in
\cite{LLM}), we think that  we can confirm it by several different methods.
Meanwhile the validity of its interpretation is much less clear.   Until the
interpretation problem is resolved in one way or another, there will remain an
open possibility  that inflationary cosmology predicts that we should live in a
center of  a spherically symmetric well.  As we will see, this possibility is
closely related to the question whether or not $\Omega = 1$ in inflationary
universe.
\vskip 0.3cm

\noindent 2.\hskip 0.4 cm It is well known that inflation makes the universe
locally flat and homogeneous. On the other hand, during each time interval
$H^{-1} = \sqrt{3\over 8\pi V(\phi)}$   the classical scalar field $\phi$
experiences quantum jumps of a typical amplitude $\delta \phi \sim {H\over
2\pi}$. (Here $H$ is the Hubble constant;  we use the system of units $M_p =
1$.) These jumps lead to small density perturbations with  almost flat spectrum
\cite{b42}.  If we follow any particular inflationary domain, we can predict
its behavior and calculate the typical amplitude of density perturbations in
it.

However, this program can be   accomplished by the methods of ref.  \cite{b42}
only if the value of the scalar field in the initial domain is small. Meanwhile
if one starts with a domain containing a sufficiently large field $\phi$,
quantum fluctuations  $\delta \phi$ become so large that they may give rise to
an eternal process of  self-reproduction of new inflationary domains. For
example, in the theory ${\lambda\over 4}\phi^4$ the self-reproduction occurs in
domains containing the field $\phi > \phi^* \sim \lambda^{-1/6}$ \cite{b19}.
Such a domain very soon becomes divided into exponentially many domains
containing all possible values of the scalar field (i.e. all possible values of
its energy density) {\it at the same moment of time}. The knowledge of initial
conditions in the original domain does not allow one to predict density
perturbations  or even mean density inside each of these new domains.  Instead
of that, one may try to study those domains which may look typical (i.e. most
abundant) in the context of the new cosmological paradigm.

In particular, one may ask the following question. Suppose that we have one
inflationary domain of initial size $H^{-1}$, containing scalar field $\phi >
\phi^*$. Let us wait 15 billion years (in synchronous time $t$ in each part of
this domain)  and see, what are the typical properties of those parts of our
original domain which at the present moment have density $10^{-29}
$g$\cdot$cm$^{-3}$. As we will show, the answer to this question proves to be
rather unexpected.

This domain exponentially expands, and becomes divided into many new domains of
size $H^{-1}$, which evolve independently of each other. In many new domains
the scalar field decreases because of classical rolling and quantum
fluctuations. The rate of expansion of these domains rapidly decreases, and
they give a relatively small contribution to the total volume of those parts of
the universe which will have density $10^{-29} $g$\cdot$cm$^{-3}$ 15 billion
years  later. Meanwhile  those domains where quantum jumps occur in the
direction of growth of the field $\phi$ gradually push this field towards the
upper bound where inflation can possibly  exist, which is presumably close to
the Planck boundary $V(\phi_p) \sim 1$. Such domains for a long time stay near
the Planck boundary,  and exponentially grow with the Planckian speed. Thus,
the longer they stay near the Planck boundary, the greater contribution to the
volume of the universe they give.

Those domains which 15 billion years later evolve into the regions with
density $10^{-29} $g$\cdot$cm$^{-3}$ cannot stay near the Planck boundary for
indefinitely long time. However, they will do their best if they stay there as
long as it is possible. In fact they will do even better  if they stay near the
Planck boundary even longer, and then rush down with the speed exceeding the
speed of classical rolling.
This may happen if quantum fluctuations  coherently add up to large quantum
jumps towards small $\phi$. This process in some sense is  dual to the process
of perpetual climbing up, which leads to the self-reproduction of inflationary
universe.

Of course, the probability of large quantum jumps down is exponentially
suppressed. However, by staying longer near the Planck boundary inflationary
domains get an exponentially large contribution to their volume. These two
exponential factors compete with each other to give us an optimal trajectory by
which the scalar field rushes down in those domains which eventually give the
leading contribution to the volume of the regions of a given density $\rho$ at
a given time $t$. From what we are saying it should be clear that the quantum
jumps of the scalar field along such optimal trajectories should have a greater
amplitude than its naively estimated value ${H\over 2\pi}$, and they should
preferably occur  in the downwards direction. As a result, the energy density
along these optimal trajectories will be smaller than the energy density of
their lazy neighbors which prefer to slide down without too much of jumping.
This creates holes in the distribution of energy density. We are going to show
 that  at any given time $t$ most of the volume of the regions of the universe
containing matter of any given density $\rho$ lies very close to the centers of
these holes.
\vskip 0.3cm

\noindent 3.\hskip 0.4 cm  The best way  to examine this scenario is to
investigate the probability distribution $P_p(\phi,t)$ to find a domain of a
given physical volume (that is why we use notation $P_p$) in a state with a
given field $\phi$ at some moment of time $t$.
 The probability distribution $P_p(\phi,t)$ obeys the following diffusion
equation (see \cite{LLM} and the references therein):
\begin{equation}\label{E372}
\frac{\partial P_p}{\partial t} =
 \frac{1}{2}  \frac{\partial }{\partial\phi}
 \left( \frac{H^{3/2}(\phi)}{2\pi}\, \frac{\partial }{\partial\phi}
\Bigl(
 \frac{H^{3/2}(\phi)}{2\pi}  P_p \Bigr) +
\frac{V'(\phi)}{3H(\phi)} \, P_p\right)  +  3H(\phi)  P_p\ .
\end{equation}
Note, that this equation is valid only during inflation, which typically occurs
within some limited interval of values of the field $\phi$: \ $\phi_{min} <
\phi < \phi_{max}$. In the simplest versions of chaotic inflation model
$\phi_{min} \sim 1$, and $\phi_{max}$ is perhaps close to the Planck boundary
$\phi_p$, where $V(\phi_p)=1$. To find solutions of this equation  one must
specify boundary
conditions. Behavior of solutions typically is not very sensitive to the
boundary conditions at the end of inflation at $\phi_{min}$; it is sufficient
to assume that the diffusion coefficient (and, correspondingly, the double
derivative term  in the r.h.s. of equation
  (\ref{E372})) vanishes for $\phi < \phi_{min}$. The
conditions near the Planck boundary $\phi = \phi_p$  play a more important
role.
In this paper we will assume, following \cite{LLM}, that there can be no
inflation at $V(\phi) > 1$, and therefore we will impose on $P_p$ the simplest
boundary condition $ P_p(\phi,t)|_{\phi>\phi_p} = 0$. At the end of the paper
we will discuss possible modifications of our results if other boundary
conditions should be imposed at the upper boundary of the region of values of
the field $\phi$ where inflation is possible.

One may try to obtain solutions of equation
(\ref{E372}) in the
form of the  series $P_p(\phi,t) =
\sum_{s=1}^{\infty} { e^{\lambda_s t}\,   \pi_s(\phi) } $. In the limit of
large time $t$ only the term with the largest eigenvalue $\lambda_1$ survives,
$P_p(\phi,t) =
e^{\lambda_1 t}\,   \pi_1(\phi) $. The function $\pi_1$ in the limit $t \to
\infty$ will have a meaning of a normalized {\it time-independent} probability
distribution to find a given field $\phi$ in a unit physical volume, whereas
the function  $e^{\lambda_1 t}$ shows the overall growth of the volume of all
parts of the universe, which does  not depend on $\phi$ in the limit $t\to
\infty$. In this limit  one can write eq. (\ref{E372}) in the form
\begin{equation} \label{eq17}
\frac{1}{2}  \frac{\partial }{\partial\phi}
 \left( \frac{H^{3/2}(\phi)}{2\pi} \frac{\partial }{\partial\phi}
\left(
 \frac{H^{3/2}(\phi)}{2\pi} \pi_1(\phi) \right) \right)
 + \frac{\partial }{\partial\phi} \left( \frac{V'(\phi)}{3H(\phi)} \,
\pi_1(\phi) \right)
 + 3H(\phi)  \cdot \pi_1(\phi) =
\lambda_1 \, \pi_1(\phi)  \ .
\end{equation}
In this paper we will concentrate on the simplest theory with $V(\phi) =
{\lambda\over 4} \phi^4$, $H = \sqrt{2\pi\lambda \over 3}\phi^2$. Eq.
(\ref{eq17}) for this theory reads \cite{LLM}:
\begin{equation}\label{r1}
\pi_1'' + \pi_1'\Bigl({6\over\lambda\phi^5} +
{9\over\phi}\Bigr) + \pi_1 \Bigl({6\over\lambda\phi^6} +
{15\over\phi^2} +
{36\pi\over\lambda\phi^4} -{\lambda_1\over \pi \phi^6}
\Bigl({6\pi\over\lambda}\Bigr)^{3/2}\Bigr) = 0\ .
\end{equation}
We have solved this equation both analytically and numerically, and have found
that
the eigenvalue $\lambda_1$ is given by $d(\lambda)\, H_{max}$, where
$d(\lambda)$ is the fractal dimension which approaches $3$ in the limit
$\lambda \to 0$, and $H_{max}$ is the maximal value of the Hubble constant in
the allowed interval of $\phi$. In our case $H_{max}= H(\phi_p) = {2\sqrt{2\pi
\over 3}}$. Thus, in the small $\lambda$ limit one has $\lambda_1 = 3 H(\phi_p)
= 2\sqrt{6\pi} \approx 8.68$.

Note, that the distribution $\pi_1$ depends on $\phi$   very sharply. For
example, one can easily check that   at small $\phi$ (at $\phi {\
\lower-1.2pt\vbox{\hbox{\rlap{$<$}\lower5pt\vbox{\hbox{$\sim$}}}}\ }
\lambda^{-1/8}\lambda_1^{-1/4}$) the leading terms in eq.  (\ref{r1}) are the
second and the last ones\footnote{The appearance  of the new scale $\phi \sim
\lambda^{-1/8}$ in the theory ${\lambda\phi^4}$ is explained in \cite{Mezh},
and was also pointed out by Mukhanov (private communication).}.  Therefore the
solution of equation (\ref{r1}) at    $\phi {\
\lower-1.2pt\vbox{\hbox{\rlap{$<$}\lower5pt\vbox{\hbox{$\sim$}}}}\ }
\lambda^{-1/8}\lambda_1^{-1/4}$  is given by
\begin{equation}\label{SMALLPHI}
\pi_1 \sim     \phi^{\sqrt{6\pi\over
\lambda}\lambda_1} \,   \  .
\end{equation}
This is an extremely strong dependence. For example, $\pi_1 \sim
\phi^{1.2\cdot 10^8}$ for the realistic value $\lambda \sim 10^{-13}$. All
surprising results we are going to obtain are rooted in this effect. One of the
 consequences is the distribution of energy density $\rho$. For example, during
inflation $\rho \approx {\lambda\over 4} \phi^4$. This implies that the
distribution of domains of  density $\rho$   is   $P_p(\rho) \sim
\rho^{3\cdot10^7}$. Thus at each moment of time $t$ the self-reproducing
inflationary universe consists of indefinitely large number of domains
containing matter with all possible values of density, the total volume of  all
domains with density $2\rho$ being approximately $10^{10^7}$ times greater than
the total volume of all domains with density $\rho$! If this result is
understood (see its discussion in \cite{LLM}), all the rest should be easy...

Let us consider now all parts of inflationary universe which contain a given
field
$\phi$ at a given moment of time $t$. One may wonder, what was the value of
this field in those domains at the moment $t - H^{-1}$ ? The  answer is simple:
One should add to $\phi$ the value of its classical drift $\dot\phi H^{-1}$.
One should also add the
amplitude of a quantum jump $\Delta \phi$. The typical   jump  is given by
$\delta \phi = \pm {H\over 2\pi}$. At the last stages  of inflation this
quantity is by many orders of magnitude smaller than $\dot\phi H^{-1}$. But in
which sense   jumps $\pm {H\over 2\pi}$ are typical? If we consider any
particular initial value of the field $\phi$, then the typical jump from this
point is indeed given by $\pm {H\over 2\pi}$. However, if we are considering
all domains with a given $\phi$ and trying to find all those domains from which
the field $\phi$ could originate back in time, the answer may be quite
different. Indeed,   the total volume of all domains with a given field $\phi$
at any moment of time $t$ strongly depends on $\phi$:   \,${P}_p(\phi) \sim
\pi_1(\phi)
\sim   \phi^{\sqrt{6\pi\over \lambda}\lambda_1} \sim \phi^{10^8}$, see eq.
(\ref{SMALLPHI}). This means that the total volume of all domains which could
jump towards the given field $\phi$ from the value $\phi +\Delta \phi$ will be
enhanced by a large  additional factor $ { {P}_p(\phi +\Delta \phi)\over
 {P}_p(\phi)} \sim   \Bigl(1+{\Delta\phi\over \phi}\Bigr)^{\sqrt{6\pi\over
\lambda}\lambda_1}$. On the other hand, the probability of large jumps
$\Delta\phi$ is suppressed by the Gaussian factor
$\exp\Bigl(-{2\pi^2\Delta\phi^2\over H^2}\Bigr)$.
One can easily verify that the product of these two factors has a sharp maximum
at $\Delta\phi = \lambda_1 \phi   \cdot {H\over 2\pi}$, and the width of this
maximum is of the order ${H\over 2\pi}$. In other words, most of the domains of
a given field $\phi$ are formed due to   jumps which are greater than the
``typical'' ones by a factor $\lambda_1 \phi \pm  1  $.
\vskip 0.3cm

\noindent 4.\hskip 0.4 cm This result is very unusual. We became partially
satisfied by our understanding of this effect only after we confirmed its
existence   by four
different methods, including computer simulations. A detailed description of
our results will be contained in a separate publication \cite{LLM2}. Here we
will briefly describe the computer simulations which we have performed.

 We have studied  a set of domains of initial size $H^{-1}$ filled with large
homogeneous field $\phi$. We considered large initial values  of $\phi$, where
the self-reproduction of inflationary domains is possible. From the point of
view of stochastic processes which we study, each domain can be modelled by a
single point with the field $\phi$ in it.  Our purpose was to study the typical
amplitude of  quantum jumps of the scalar field $\phi$ in those domains which
reached some value $\phi_0$ close to the end of inflation.

Each step of our calculations corresponds to a time change
$\Delta t = u H^{-1}_0$. Here $H_0 \equiv H(\phi_0)$,  and
$u$ is some number, $u < 1$.  The results do not depend on   $u$ if  it is
  small enough.
The evolution of the field $\phi$ in each domain consists of  several
independent parts.  First of all, the field evolves according
to classical equations of motion during inflation. Secondly,
it makes quantum jumps by $\delta\phi ={H  \over \pi}
\sqrt{u H \over 2H_0}\,  \sin r _i$. Here $r_i $ is a
set of random numbers,  which are different for each inflationary domain.
To account for the growth of physical volume of each domain we   used  the
following procedure. We followed each domain until its radius grows two times,
and after
that we considered it as 8 independent
domains.  If  we would continue doing so
for a long time, the number of such domains (and our
distribution $P_p(\phi,t)$) would become exponentially large. We will describe
in \cite{LLM2} the procedure which we used to overcome this difficulty. In
accordance with  our condition $ P_p(\phi,t)|_{\phi>\phi_p} = 0$, we removed
all domains where the field $\phi$ jumped to the super-Planckian densities
$V(\phi) > 1$.
After a sufficiently large time $t$ the distribution of domains followed by the
computer with a good accuracy approached the stationary distribution $\pi_1$
which we have obtained in \cite{LLM} by a completely different method. We used
it as a consistency check for our calculations.

We kept in the computer memory information about all jumps of all domains
during the last time interval $H_0^{-1}$. This made it possible to evaluate an
average sum of all jumps of those domains in which the scalar field  became
smaller than some value $\phi$ within the last time interval $H_0^{-1}$.
Naively, one could expect   this value to be smaller than ${H_0\over 2\pi}$,
since the average amplitude of the jump is ${H_0\over 2\pi}$, but they occur
both in the positive and negative directions. However,  our simulations
confirmed our analytical result $\Delta\phi = \lambda_1 \phi   \cdot {H_0\over
2\pi}$. In other words, we have found  that most of the domains which   reach
the hypersurface  $\phi = \phi_0$ within a time interval $\Delta t = H_0^{-1}$
do it by rolling accompanied by persistent jumps down, which have a combined
amplitude $\lambda_1 \phi_0 $ times greater than ${H_0\over 2\pi}$.

Even though we obtained this result in the context of the theory ${\lambda\over
4} \phi^4$, it can be easily generalized. One can show that for any
inflationary theory in the regime where one can neglect the first (diffusion)
term  in the l.h.s. of equation
  (\ref{eq17}) (which is usually possible at the last stages  of inflation),
most of the domains of a given field $\phi$ are formed due to   jumps which are
greater than the
``typical'' ones by a factor $\lambda_1 {4 V(\phi)\over V'(\phi)} $
\cite{LLM2}.
\vskip 0.3cm

\noindent 5.\hskip 0.4 cm As we already
mentioned, the probability of large fluctuations should be suppressed  by the
factor
$\exp\Bigl(-{2\pi^2\delta\phi^2\over H^2}\Bigr)$, which in our case gives the
suppression factor $ \sim  \exp(-10^3)$. It is well known that exponentially
suppressed perturbations typically give rise to spherically symmetric bubbles
\cite{Adler}.
Note also, that the Gaussian distribution suppressing the amplitude of the
perturbations refers to the amplitude of a perturbation in its maximum. Let us
show now that the main part of the volume of the universe in a state with a
given $\phi$ (or with a given density $\rho$) corresponds to the centers of
these bubbles.

Consider again the part of the universe with   a given $\phi$ at a given time
$t$. We have found that most of the jumps producing this field $\phi$ during
the previous time interval $H^{-1}$ occurred from a very narrow region $\phi
+{\dot \phi\over H} +\lambda_1 \phi {H\over 2\pi}$ of values of the scalar
field.  The width of this region was found to be of the order of ${H\over
2\pi}$, which is much smaller than the typical depth of our bubble $\Delta\phi
\sim \lambda_1 \phi {H\over 2\pi}$. Now suppose that the domain containing the
field $\phi$ appears not at the center of the bubble, but at its wall. This
would mean that the field  near the center of the bubble is somewhat smaller
than $\phi$. Such a configuration could be created by a jump  from  $\phi
+{\dot \phi\over H} + \lambda_1 \phi {H\over 2\pi}$ only if the amplitude of
the jump is  greater   than  $\lambda_1 \phi {H\over 2\pi}$.  However, we have
found that the main contribution to the volume of domains with a given $\phi$
is produced by jumps of an amplitude  $\lambda_1 \phi{H\over 2\pi}   \pm
{H\over 2\pi}$, the greater deviation from the typical amplitude $\lambda_1
\phi {H\over 2\pi}$ being exponentially suppressed.  This means that the scalar
field $\phi$ can differ from its value at the center of the bubble by no more
than the usual amplitude of scalar field perturbations ${H\over 2\pi}$, which
is smaller than the depth of the bubble by a factor $(\lambda_1\phi)^{-1}$.
Thus, the main fraction of the volume of the universe with a given $\phi$ (or
with a given density of matter) can be only slightly outside the center, which
may lead to a small dipole anisotropy of the microwave background radiation.

We should emphasize that our results are based on the investigation of  the
global structure of the universe rather than of the structure of each
particular bubble. If one neglects that the universe is a fractal and looks
only at one particular bubble (i.e. at the one in which we live now), then one
can easily see that inside  each bubble there is a plenty of space far away
from its center. Therefore one could conclude that there is nothing special
about the centers of the bubbles. However,  when determining the fraction of
domains near the centers we were   comparing   the volumes of {\it all} regions
of {\it equal} density at equal time. Meanwhile, the density $\rho_{\rm wall}$
of matter on the walls of a bubble is greater than the density $\rho_{\rm
center}$ in its center. As we have emphasized in the discussion after eq.
(\ref{SMALLPHI}),   the total volume of {\it all} domains of density $\rho_{\rm
wall}$ is greater than the total volume of all domains of  density $\rho_{\rm
center}$ by the   factor $({\rho_{\rm wall}/\rho_{\rm center}})^{3\cdot 10^7}$.
 Thus, the volume of space outside the center of the bubble is not smaller than
the space near the center. However, going outside the center brings us to the
region of a different  density,  $ \rho_{\rm wall} > \rho_{\rm center}$.  Our
results imply that one can find much more space with $\rho = \rho_{\rm wall}$
not at the walls of our bubble, but near  the centers of {\it other} bubbles.
\vskip 0.3cm

\noindent 6.\hskip 0.4 cm  The nonperturbative jumps down should occur on all
scales independently. At the earlier stage, when the time changed from $t-
2H^{-1}$ to $t- H^{-1}$, the leading contribution to the volume of the universe
of a given density should be also given by domains which are jumping down by
$\Delta \phi \sim  \lambda_1 \phi\, {H\over 2\pi}$. This implies that on each
new length scale different from the previous one by the factor of $e$ (one new
e-folding) our trajectory runs down creating a local depression of the scalar
field equal to $\lambda_1 \phi\, {H\over 2\pi}$. One may visualize the
resulting distribution of the scalar field in the following way.
At some scale $r$ the deviation of the field $\phi$ from homogeneity can be
approximately represented as a hole of a radius $r$ with the depth $\lambda_1
\phi\, {H\over 2\pi}$. Near the bottom of this hole there is another hole of a
smaller radius   $e^{-1}r$ and approximately of the same depth $\lambda_1
\phi\, {H\over 2\pi}$. Near the center of this hole there is another hole of a
radius $ e^{-2}r$, etc. Of course, this is just a discrete model. The shape of
the smooth distribution of the scalar field is determined by  the equation
\begin{equation}\label{ScalarShape1}
{d\phi\over d\ln rH} =   \lambda_1 \phi\, {H\over 2\pi} =   \sqrt{ \lambda
\over 6 \pi}\, \lambda_1   \phi^3 \,  ,
\end{equation}
which gives
\begin{equation}\label{ScalarShape2}
\phi ^2(r) \approx  {\phi ^2(0) \over  1 -  \lambda_1 \, \phi^{2}(0) \sqrt{
2\lambda\over 3\pi }\,      \ln {rH}}   ~~~~~~  \, \mbox{for} \ \  r > H^{-1} \
{}.
\end{equation}
Note that   $\phi(r) \approx \phi(0)$  for $r < H^{-1}$ (there are no
perturbations of  the classical field on this scale).

This  distribution is altered by the usual small perturbations of the scalar
field. These perturbations are responsible for the large nonperturbative jumps
down when they add up  near the center of the hole, but at a distance much
greater than their wavelength from the center of the hole these perturbations
have the usual magnitude ${H\over 2\pi}$.  Thus, our results do not lead to
considerable modifications of the usual density perturbations which lead to
galaxy formation. However, the presence of the deep hole (\ref{ScalarShape2})
can significantly change the local geometry of the universe.

 In the inflationary scenario with $V(\phi) =
{\lambda\over 4} \phi^4$ fluctuations which presently have the scale comparable
with the horizon radius $r_h \sim 10^{28}$ cm have been formed at $\phi~\sim~5$
(in the units $M_p = 1$).
As we have mentioned already $\lambda_1 \approx   2\sqrt{6\pi} \sim 8.68$ for
our choice of
boundary conditions \cite{LLM}. This means that the  jump down on the scale of
the present horizon should be    $\lambda_1 \phi \sim 40$
times greater than the standard jump. In the theory ${\lambda\over 4} \phi^4$
the standard jumps   lead to   density
perturbations of the   amplitude ${\delta\rho\over  \rho} \sim
{2\sqrt{6\lambda\pi}\over 5}\, \phi^3  \sim 5\cdot10^{-5}$ (in the
normalization of \cite{MyBook}). Thus, according to our  analysis, the
nonperturbative decrease of density on each length scale different from the
previous one by the factor $e$  should be about ${\delta\rho\over  \rho} \sim
\lambda_1 {2\sqrt{6\lambda\pi}\over 5} \phi^4 \sim 2\cdot10^{-3}$.  This allows
one to evaluate the shape of the resulting hole in the density distribution as
a function of the distance from its center. One can write the following
equation for the scale dependence of density:
\begin{equation}\label{shape1}
{1\over \rho} {d\rho\over d\ln {r\over r_0}} = -   \lambda_1   \cdot
{2\sqrt{6\lambda\pi}\over 5}\, \phi^4 \ ,
\end{equation}
where $r$ is the distance from the center of the hole. Note that $\phi =
{1\over\sqrt \pi}\,(\ln {r\over r_0})^{1\over 2}$ in the theory ${\lambda \over
4}\phi^4$ \cite{MyBook}. Here $ r_0$ corresponds to the smallest scale at which
inflationary perturbations have been produced. This scale is model-dependent,
but typically at present it is about 1  cm. This yields
\begin{equation}\label{shape2}
{\Delta\rho\over \rho_c} \equiv {\rho(r) - \rho(r_0)\over \rho(r_0)} =    {2
\lambda_1\sqrt{2\lambda}\over 5\pi\sqrt{3\pi}} \,   \ln^3{r\over r_0}    \  .
\end{equation}
This gives   the typical deviation of the density on the scale of the horizon
(where $\ln{r_h\over r_0}\sim 60$) from the density at the center:
${\Delta\rho\over \rho_c} \sim 750   \cdot {\delta\rho\over \rho}
\sim 4\cdot 10^{-2}$.
\vskip 0.3cm

\noindent 7.\hskip 0.4 cm It is very tempting to interpret this effect in such
a way that the universe
around us becomes locally open, with $1 - \Omega  \sim  10^{-1}$. Indeed, our
effect is very similar to the one discussed in \cite{ColemanDL,Turok}, where it
was shown that the universe becomes open if it is contained in the interior of
a bubble created by the $O(4)$ symmetric tunneling.  Our nonperturbative  jumps
  look very similar to tunneling with  the  bubble formation.  However, unlike
in the case considered in \cite{ColemanDL,Turok}, our bubbles appear on all
length scales. This removes the problem of fine-tuning, which plagues the
one-bubble models of the universe \cite{Turok}. However, it rises another
problem, which may be even more complicated.

 Note,  that the results discussed above refer to the density distribution at
the moment when the corresponding wavelengths were   entering horizon. At the
later stages  gravitational instability should lead to growth of the
corresponding density perturbations. Indeed, we know that density perturbations
on the galaxy scale have grown more than $10^4$ times in the linear growth
regime until they reached the amplitude ${\delta\rho\over \rho} \sim 1$, and
then continued growing even further. The same can be expected in our case, but
even in a more dramatic way since  our ``density perturbations"  on all scales
are about 40 times greater than the usual density perturbations which are
responsible for galaxy formation. This would make the center of the hole  very
deep; its density should be many orders of magnitude smaller than the density
of the universe on the scale of horizon. Moreover, the center would be devoid
of any structures necessary for the existence of our life. Indeed, on each
particular scale the jump down completely overwhelms the amplitude of usual
density perturbations. The  bubble  cannot contain any galaxies at the distance
from the center comparable with the galaxy scale, it cannot contain any
clusters at the distance comparable with the size of a cluster, etc.

This problem can be easily resolved. Indeed, our effect (but not the amplitude
of the usual  density perturbations) is proportional to $\lambda_1$, which is
determined by the maximal value of the Hubble constant compatible with
inflation. If, for example, the maximal energy scale in quantum gravity or in
string theory is given not by $10^{19}$ GeV, but by $10^{18}$ GeV, then the
parameter $\lambda_1$ will decrease by a factor $10^{-2}$. Other ways to change
$\lambda_1$ are described in \cite{GBLL}. Thus it is easy to make our effect
very small  without disturbing the standard predictions of inflationary
cosmology.

It is possible though that we will not have any problems even with a large
$\lambda_1$ if we  interpret our results more carefully.  An implicit
hypothesis behind our interpretation is that we are typical, and therefore we
live and make observations in those parts of the universe where  most other
people do.  One may argue that the total number of observers which can live in
domains with given properties (e.g. in domains with a given density) should be
proportional to the total volume of these domains at a given time. However, our
existence is  determined not by the local density of the universe  but  by the
possibility for life to evolve for about  5 billion years on a planet of our
type in a vicinity of a star of the type of the Sun. If, for example, we have
density $10^{-29} $g$\cdot$cm$^{-3}$ inside the center of the hole, and density
$10^{-27} $g$\cdot$cm$^{-3}$ on the horizon, then the age of our part of the
universe (or, to be more accurate, the time after the end of inflation) will be
determined by the average density  $10^{-27} $g$\cdot$cm$^{-3}$, and it will be
smaller than 5 billion years.  Moreover, as we have mentioned above, any
structures such as galaxies or clusters cannot be formed near the centers of
the holes.

Thus, the naive idea that the number of observers is proportional to volume
does not work  at the distances from the centers which are much smaller than
the present size of the horizon. Even though at any given moment of time most
of the volume of the universe at the density $10^{-29} $g$\cdot$cm$^{-3}$ is
concentrated near these holes, the corresponding parts of the universe are too
young and do not have any structures necessary for our existence. Volume alone
does not mean much. We live on the surface of the Earth even though the volume
of empty space around us is incomparably greater.

On the other hand, the naive relation between the number of observers and the
volume in a state of a given density may work nicely for the matter
distribution on the scale of the present horizon. These considerations may make
the distribution (\ref{shape2}) homogeneous near the center not up to the scale
$r_0 \sim1$
cm,  but up to some much greater scale   $r \gg 10^{22}$ cm. The resulting
distribution may appear relatively
smooth. It may resemble an open universe with a scale-dependent  effective
parameter $\Omega(r)$.

 An additional ambiguity in our interpretation appears due to the dependence of
the distribution $P_p$ on the choice of time parametrization. Indeed,  there
are many different ways to define ``time'' in general relativity. If, for
example, one measures time not by clock but by rulers and determines time  by
the degree of a local expansion of the universe, then in this ``time'' the rate
of expansion of the universe does not depend on its density. As a result,   our
effect is absent in this time parametrization \cite{LLM}. In this paper we used
 the standard time parametrization which is most closely related to our  own
nature (time  measured by number of oscillations rather than  by the distance
to the nearby galaxies). But maybe we should use another time parametrization,
or even integrate over all possible time parametrizations? The last
possibility, suggested to us by Robert Wagoner, is very interesting, since
after the integration we would obtain a parametrization--independent measure in
quantum cosmology. Even more radical possibility is the integration over all
possible values of time. However,  the corresponding integral diverges in the
limit $t \to \infty$ \cite{GBLL}. Right now we still do not know  what is   the
right way to go. We do not even know if it is right that we are typical and
that we should live in domains of the greatest volume, see the discussion of
this problem in \cite{LLM,GBLL}.

Until  the interpretation problem is resolved, it will remain unclear whether
our result  is just a mathematical curiosity similar to the twin paradox, or it
can be considered as a  real prediction of properties of our part of the
universe. On the
other hand,  at present we cannot exclude the last possibility, and this by
itself  is a very unexpected conclusion. Few years ago we would say that the
possibility that we live in a local ``center of the world'' definitely
contradicts   inflationary cosmology. Now we can only say that  it is an open
question to be studied both theoretically and experimentally. As we have seen,
the relation of our result to the properties of our part of the universe
depends on   interpretation of quantum cosmology, and the amplitude of our
effect is very sensitive to the properties of the universe at nearly Planckian
densities. This suggests that unexpected (from the point of view of inflation)
observational data, such as possible local deviations from $\Omega = 1$, or
possible dependence of the Hubble constant on the length scale, may tell us
something important about quantum cosmology and  particle physics   near the
Planck  density.
\vskip 0.4cm

The authors are very grateful to  J. Garc\'{\i}a--Bellido, L. Kofman, V.
Mukhanov, A. Starobinsky and R. Wagoner for many enlightening discussions.
This work was
supported in part  by NSF grant PHY-8612280.


\vfill

\begin{thebibliography}{999}
\bibitem{LLM} A.D. Linde and A. Mezhlumian,  {\it Phys.\   Lett.}
{\bf B307}   (1993)  25;\\   A.D. Linde, D.A. Linde  and
A. Mezhlumian,  {\it Phys.\   Rev.} {\bf D49}  (1994)  1783.
\bibitem{b42} V.F. Mukhanov and G.V. Chibisov, JETP Lett. {\bf 33}  (1981) 523;
S.W. Hawking, Phys. Lett. {\bf 115B}  (1982) 339;
A.A. Starobinsky, Phys. Lett. {\bf 117B}  (1982)  175;
A.H. Guth and S.-Y. Pi, Phys. Lett. {\bf 49}  (1982) 1110;
J. Bardeen, P.J. Steinhardt and M. Turner, Phys. Rev. {\bf D28}  (1983)   679;
V.F. Mukhanov, JETP Lett. {\bf 41}  (1985)  493.
\bibitem{b19} A.D. Linde, Phys. Lett. {\bf 175B}  (1986) 395.
\bibitem{Mezh} A. Mezhlumian, ``The Branching Universe'', Stanford University
preprint, to appear.
\bibitem{LLM2}  A.D. Linde, D.A. Linde  and
A. Mezhlumian,  in preparation.
\bibitem{Adler} R.J. Adler, {\it The Geometry of Random Fields} (Wiley,
Chichester, 1981);\\
 J.M. Bardeen, J.R. Bond, N. Kaiser and A.S. Szalay, Astrophys. J. {\bf 304}
(1986) 15.
\bibitem{MyBook} A.D. Linde, {\it Particle Physics and Inflationary
Cosmology} (Harwood, Chur, Switzerland, 1990).
\bibitem{ColemanDL} S. Coleman and  F. De Luccia,  Phys. Rev.  {\bf D21} (1980)
3305; J.R. Gott,  Nature {\bf 295}
(1982) 304;  A.H. Guth and  E.J. Weinberg, Nucl. Phys. {\bf B212} (1983) 321;
M. Sasaki, T. Tanaka, K.  Yamamoto, and J. Yokoyama, Phys. Lett. {\bf B317}
(1993) 510.
\bibitem{Turok} M. Bucher,  A.S. Goldhaber, and N. Turok,  ``An Open Universe
{}From Inflation'', Princeton University preprint  PUPT {1507},
 hep-ph/9411206 (1994).
\bibitem{GBLL} J. Garc\'{\i}a--Bellido, A.D. Linde and D.A.
Linde, {\it Phys. Rev.} {\bf D50}  (1994) 730;\\
 J. Garc\'{\i}a--Bellido and
A.D. Linde, ``Stationarity of Inflation and Predictions of Quantum Cosmology'',
Stanford  University preprint SU-ITP-94-24, hep-th/9408023, to be published in
Phys. Rev.
D.


\end{thebibliography}
\end{document}